# Information leakage resistant quantum dialogue against collective noise

Tian-Yu Ye*

College of Information & Electronic Engineering, Zhejiang Gongshang University, Hangzhou 310018, P.R.China

In this paper, two information leakage resistant quantum dialogue (QD) protocols over a collective-noise channel are proposed. Decoherence-free subspace (DFS) is used to erase the influence from two kinds of collective noise, i.e., collective-dephasing noise and collective-rotation noise, where each logical qubit is composed of two physical qubits and free from noise. In each of the two proposed protocols, the secret messages are encoded on the initial logical qubits via two composite unitary operations. Moreover, the single-photon measurements rather than the Bell-state measurements or the more complicated measurements are needed for decoding, making the two proposed protocols easier to implement. The initial state of each logical qubit is privately shared between the two authenticated users through the direct transmission of its auxiliary counterpart. Consequently, the information leakage problem is avoided in the two proposed protocols. Moreover, the detailed security analysis also shows that Eve's several famous active attacks can be effectively overcome, such as the Trojan horse attack, the intercept-resend attack, the measure-resend attack, the entangle-measure attack and the correlation-elicitation (CE) attack.

**Quantum dialogue(QD); bidirectional quantum secure direct communication(BQSDC); information leakage; collective-dephasing noise; collective-rotation noise**

## 1 Introduction

In 1984, the principle of quantum mechanics was introduced into the classical communication, giving birth to a brand new area subsequently called quantum secret communication, when Bennett and Brassard proposed the first quantum key distribution (QKD) protocol (i.e., the BB84 protocol[1]). Since then, many works on QKD[2-5] have been presented. It is well known that the goal of QKD is to establish an unconditionally secure key between two remote authenticated users. In order to realize the direct transmission of confidential messages between two remote authenticated users without establishing a prior key to encrypt and decrypt them in advance, in 2000, Long and Liu[6] put forward the first quantum secure direct communication (QSDC) protocol, i.e., the efficient two-step QSDC protocol (its simplified version is then published in 2002). In this protocol, they proposed the concept of quantum data block and the idea of security check based on quantum data block for the first time. In 2002, Boström and Felbinger[7] put forward the famous ping-pong protocol using Einstein-Podolsky-Rosen (EPR) pairs as quantum information carriers. In 2003, Deng et al. [8] put forward the two-step QSDC protocol with EPR pairs using quantum data block transmission and unitary operation encoding. In this protocol, they suggested the physical model for quantum secure direct communication, and analyzed the security check method in detail. Moreover, they discussed the case of a noisy channel and the experimental implementation, and gave the requirements for quantum secure direct communication. In 2004, Cai and Li[9] doubled the capacity of ping-pong protocol by introducing two additional unitary operations; they also proposed a deterministic direct communication protocol using single qubit in a mixed state[10]. It should be emphasized that all the QSDC protocols in Refs.[7,9-10] are actually quasi-secure. In 2004, Deng and Long[11] proposed the QSDC protocol based on single photons for the first time, i.e., the four-state two-way quantum communication protocol, which is feasible both in theory and under experimental conditions. In this protocol, the basic requirements for constructing quantum secure direct communication are definitely pointed out. In 2005, Wang et al.[12] proposed a QSDC protocol with quantum superdense coding, which is an important application of high-dimension quantum system; Wang et al.[13] also proposed a multi-step QSDC protocol using multi-particle Green-Horne-Zeilinger (GHZ) state. In 2007, Li et al.[14] put forward a QSDC protocol with quantum encryption using pure entangled states. In 2008, Chen et al.[15] proposed a novel controlled QSDC protocol with quantum encryption using a partially entangled GHZ state; Chen et al.[16] also proposed a novel three-party controlled QSDC protocol based on W state. In 2011, Wang et al.[17] proposed a high-capacity QSDC protocol based on quantum hyperdense coding with hyperentanglement; Gu et al.[18] put forward a bidirectional QSDC network protocol with hyperentanglement; Gu et al.[19] also put forward a two-step QSDC protocol with hyperentanglement in both the spatial-mode and the polarization degrees of freedom of photon pairs; Gu et al.[20] also put forward a kind of robust QSDC protocol with a quantum one-time pad over a collective-noise channel; Shi et al.[21] proposed a kind of QSDC protocol using three-dimensional hyperentanglement; Gao et al.[22] put forward a high-capacity QSDC protocol by swapping entanglements of 3x3-dimensional Bell states. In 2012, Sun et al.[23] proposed a QSDC protocol using two-photon four-qubit cluster states; Liu et al.[24] put forward a high-capacity QSDC protocol with single photons in both the polarization and the spatial-mode degrees of freedom. In 2013, Tsai and Hwang[25] proposed a deterministic quantum communication protocol using the symmetric W state; Ren et al.[26] put forward a robust two-step QSDC protocol based on the spatial-mode Bell states and the photonic spatial Bell-state analysis.

Although QSDC is able to realize the direct transmission of confidential messages from one authenticated user to the other one, it still has an apparent drawback. That is, QSDC can not realize the mutual exchange of secret messages between two remote authenticated users. In order to get rid of this drawback, the novel concept of bidirectional quantum secure direct communication (BQSDC) was proposed by Zhang et al.[27-28] and Nguyen[29] in 2004, which is also called quantum dialogue (QD) in the

*Corresponding author:

E-mail：happyyty@aliyun.com

literature. Since then, QD has made rapid progress [30-39]. However, the phenomenon of classical correlation or information leakage in QD was discovered by Tan and Cai[40] and Gao et al.[41-42] in 2008, which means that QD may run a great security risk. Unfortunately, information leakage exists in all QD protocols of Refs.[27-39]. Since then, how to solve the information leakage problem in QD has quickly become an active research topic. As a result, several information leakage resistant QD protocols[43-48] were constructed from different ways of implementation. For example, in Ref.[43], Shi et al. put forward a QD protocol without information leakage based on a shared private Bell state; in Ref.[44], Shi et al. put forward a QD protocol without information leakage based on a shared private single photon; in Ref.[45], Shi proposed a QD protocol without information leakage based on the correlation extractability of Bell state and the auxiliary single particle; in Ref.[46], Gao suggested two QD protocols without information leakage based on the measurement correlation from the entanglement swapping between two Bell states; in Ref.[47], the author proposed a large payload QD protocol without information leakage based on the entanglement swapping between any two GHZ states and the auxiliary GHZ state; in Ref.[48], in cooperation with Jiang, the author proposed a QD protocol without information leakage based on the entanglement swapping between any two Bell states and the auxiliary Bell state. It is apparent that in order to overcome the information leakage problem, all the QD protocols in Refs.[43-44,47-48] use the direct transmission of auxiliary quantum state to realize the initial quantum state sharing between two authenticated users. However, there exists a common weak point in the above information leakage resistant QD protocols[43-48]. That is, they are merely suitable for an ideal environment, which means that they can not work in a noisy environment. As we know, in a practical application, a quantum communication protocol inevitably faces the disturbance from noise. But the study has not begun on the information leakage resistant QD protocol with an anti-noise property. Therefore, constructing the information leakage resistant QD protocol with an anti-noise property is an emergent task at present.

In the literature of quantum secret communication, since they always weakly interact with the environments, photons are always regarded as the best candidates for information carriers. However, due to the thermal fluctuation, vibration and the imperfection of fiber, photons are inevitably disturbed by the noise during the practical transmission via a fiber. At present, several good methods have been suggested to erase the influence from noise, such as entanglement purification [49], quantum error correct code (QECC)[50], single-photon error rejection (SPER)[51] and decoherence-free subspace (DFS)[20,52-59]. However, each of entanglement purification, QECC and SPER has a distinct drawback[20,56-57]:entanglement purification always needs to consume infinite quantum resources while distilling the perfect maximally entangled states from a mixed ensemble; QECC always encodes one logical bit into several physical qubits according to the type of noise, and then the user measures the stabilizer codes to detect and correct the errors; the SPER scheme always succeeds probabilistically although it needs less quantum resources. As a result, DFS has been popularly used to construct the quantum secret communication protocols with an anti-noise property, as the logical qubits in DFS always suffer from the same noise and are invariant towards it. For example, by using DFS, Yang and Hwang [59] proposed two pioneering anti-noise QD protocols without information leakage in 2013.

Based on the above analysis, in this paper, two information leakage resistant QD protocols over a collective-noise channel are proposed. DFS is used to erase the influence from two kinds of collective noise, i.e., collective-dephasing noise and collective-rotation noise, where each logical qubit is composed of two physical qubits and free from noise. In each of the two proposed protocols, the secret messages are encoded on the initial logical qubits via two composite unitary operations. Moreover, the single-photon measurements rather than the Bell-state measurements or the more complicated measurements are needed for decoding, making the two proposed protocols easier to implement. The initial state of each logical qubit is privately shared between the two authenticated users through the direct transmission of its auxiliary counterpart. Consequently, the information leakage problem is avoided in the two proposed protocols. Moreover, the detailed security analysis also shows that Eve's several famous active attacks can be effectively overcome, such as the Trojan horse attack, the intercept-resend attack, the measure-resend attack, the entangle-measure attack and the correlation-elicitation (CE) attack.

## 2 Description of QD protocols
### 2.1 Information leakage resistant QD protocol against collective-dephasing noise

The collective-dephasing noise in a quantum channel can be modeled as [20,52,54-59]

$$U_{dp}|H\rangle = |H\rangle,$$
$$U_{dp}|V\rangle = e^{i\varphi}|V\rangle. \qquad (1)$$

where $\varphi$ is the parameter of collective-dephasing noise which fluctuates with time. The vectors $|H\rangle$ and $|V\rangle$ are the horizontal and the vertical polarizations of photons, respectively. Apparently, the logical qubit composed of two physical qubits with an antiparallel parity is immune to this kind of noise. Two logical qubits with this property are [20,52,54-58]

$$\left|0_{dp}\right\rangle_L = |H\rangle_{A_1}|V\rangle_{A_2}, \quad \left|1_{dp}\right\rangle_L = |V\rangle_{A_1}|H\rangle_{A_2}. \qquad (2)$$

The superpositions of these two logical qubits, i.e., [20,54-55,58-59]

$$\left|\pm x_{dp}\right\rangle_L = \frac{1}{\sqrt{2}}\left(\left|0_{dp}\right\rangle_L \pm \left|1_{dp}\right\rangle_L\right) = \frac{1}{\sqrt{2}}\left(|H\rangle_{A_1}|V\rangle_{A_2} \pm |V\rangle_{A_1}|H\rangle_{A_2}\right) \equiv \left|\psi^\pm\right\rangle_{A_1A_2},$$ are also immune to this kind of noise. Thus, the four

states $|0_{dp}\rangle_L, |1_{dp}\rangle_L, |+x_{dp}\rangle_L$ and $|-x_{dp}\rangle_L$ span a DFS on this kind of noise. Obviously, $Z_L^{dp} = \{|0_{dp}\rangle_L, |1_{dp}\rangle_L\}$ and $X_L^{dp} = \{|+x_{dp}\rangle_L, |-x_{dp}\rangle_L\}$ are two corresponding measuring bases for the above four states, respectively. Define two composite unitary operations as

$$U^{dp_0} = I_{A_1} \otimes I_{A_2}, \quad U^{dp_1} = (-i\sigma_y)_{A_1} \otimes (\sigma_x)_{A_2}, \tag{3}$$

where $I = |H\rangle\langle H| + |V\rangle\langle V|$, $-i\sigma_y = |V\rangle\langle H| - |H\rangle\langle V|$ and $\sigma_x = |V\rangle\langle H| + |H\rangle\langle V|$. It is clear that the following relations exist: [20]

$$U^{dp_1}|0_{dp}\rangle_L = |1_{dp}\rangle_L, \quad U^{dp_1}|1_{dp}\rangle_L = -|0_{dp}\rangle_L, \quad U^{dp_1}|+x_{dp}\rangle_L = -|-x_{dp}\rangle_L, \quad U^{dp_1}|-x_{dp}\rangle_L = |+x_{dp}\rangle_L. \tag{4}$$

The composite unitary operation $U^{dp_1}$ can not alter each of the measuring bases itself but flip its two states inside it. Note that the counterpart of this property in an ideal condition was previously described in Deng and Long's one-time pad QSDC[11].

Suppose that Alice has $N$ bits secret messages $\{k_1, k_2, \cdots, k_N\}$ and Bob has $N$ bits secret messages $\{i_1, i_2, \cdots, i_N\}$, where $k_n, i_n \in \{0,1\}, n \in \{1, 2, \cdots, N\}$. Let each of $U^{dp_0}$ and $U^{dp_1}$ represent one-bit secret message in such a way that $\{U^{dp_0} \to 0, U^{dp_1} \to 1\}$. The implementation steps of information leakage resistant QD protocol against collective-dephasing noise can be depicted as follows.

**Step 1: Bob's Preparation and transmission.** Bob prepares a sequence of $2N$ logical qubits, i.e., $S = \{L_1, L_1', L_2, L_2', \cdots, L_n, L_n', \cdots, L_N, L_N'\}$, making each two adjacent logical qubits $L_n$ and $L_n'$ $(n = 1, 2, \cdots, N)$ in the same state (randomly in one of the four states $\{|0_{dp}\rangle_L, |1_{dp}\rangle_L, |+x_{dp}\rangle_L, |-x_{dp}\rangle_L\}$). Similar to Long and Liu's protocol[6], quantum data block transmission is adopted to transmit the prepared logical qubits from Bob to Alice. For the sake of security, Bob uses the decoy photon technique[60-61] to check whether the quantum channel is safe or not. That is, Bob prepares $\delta_1 + \delta_2$ logical qubits randomly in one of the above four states as the decoy logical photons, and randomly inserts these decoy logical qubits into sequence $S$. As a result, a new sequence $S'$ forms. Finally, Bob lets the two photons in each logical qubit from sequence $S'$ pass through an optical fiber with a time window shorter than the variation of noise,[20,56] and sends them to Alice.

**Step 2: The first security check.** After Alice announces Bob the receipt of sequence $S'$, they start to implement the first security check: (1) Bob tells Alice the positions and the preparation bases of $\delta_1$ decoy logical qubits; (2) Alice measures $\delta_1$ decoy logical qubits using Bob's preparing bases and tells Bob her measurement results; (3) Bob judges whether the quantum channel is secure or not by comparing the initial states of $\delta_1$ decoy logical qubits with Alice's measurement results. If there is no eavesdropping, the communication is continued from the next step. Otherwise, the communication is halted.

**Step 3: Alice's encoding and transmission.** Alice discards $\delta_1$ decoy logical qubits in sequence $S'$. Bob tells Alice the positions of $\delta_2$ decoy logical qubits. Alice picks up $\delta_2$ decoy logical qubits, and stores the remaining $2N$ logical qubits to recover sequence $S$. Alice divides sequence $S$ into $N$ message groups in the way that each two adjacent logical qubits $L_n$ and $L_n'$ $(n = 1, 2, \cdots, N)$ forms a group. Alice and Bob agree that only the first logical qubit in each message group is used for encoding. Then, Alice encodes her one-bit secret message $k_n$ by performing the composite unitary operation $U_n^{dp_{k_n}}$ on the logical qubit $L_n$ from the $n^{th}$ message group. Accordingly, the logical qubit $L_n$ is changed into $U_n^{dp_{k_n}} L_n$. Consequently, the $n^{th}$ message group is turned into $(U_n^{dp_{k_n}} L_n, L_n')$. Then, Alice takes the first logical qubit out from each message group to form a new sequence $L$. That is, $L = \{U_1^{dp_{k_1}} L_1, U_2^{dp_{k_2}} L_2, \cdots, U_n^{dp_{k_n}} L_n, \cdots, U_N^{dp_{k_N}} L_N\}$. The remaining logical qubit from each message group forms a new sequence $L'$, i.e., $L' = \{L_1', L_2', \cdots, L_n', \cdots, L_N'\}$. Alice also encodes her checking message on each of $\delta_2$ decoy logical qubits by performing one of the two composite unitary operations $U^{dp_0}$ and $U^{dp_1}$. Then, Alice randomly inserts these $\delta_2$ encoded decoy logical qubits into sequence $L$. As a result, a new sequence $L''$ is derived. Finally, Alice sends sequence $L''$ to Bob, and keeps $L'$ by herself.

**Step 4: The second security check.** After Bob announces Alice the receipt of sequence $L''$, they start to implement the second security check: (1) Alice tells Bob the positions of the $\delta_2$ encoded decoy logical qubits in sequence $L''$; (2) Since he prepares the $\delta_2$ decoy logical qubits by himself, Bob can know their initial states and the measuring bases of the $\delta_2$ encoded decoy logical qubits. Bob selects the right measuring bases to measure the $\delta_2$ encoded decoy logical qubits to decode Alice's checking messages. Afterward, Bob announces Alice his decoding results about her checking messages; (3) Alice compares her checking messages with Bob's announcement. If there is no eavesdropping, the communication is continued from the next step. Otherwise, the communication is halted.

**Step5: Bob's encoding and their decoding.** Bob discards the $\delta_2$ encoded decoy logical qubits in sequence $L^{''}$. As a result, sequence $L^{''}$ is turned into sequence $L$. Then, Bob encodes his one-bit secret message $i_n$ by performing the composite unitary operation $U_n^{dp_{i_n}}$ on the logical qubit $U_n^{dp_{k_n}} L_n$. Accordingly, the logical qubit $U_n^{dp_{k_n}} L_n$ is changed into a new logical qubit $U_n^{dp_{i_n}} U_n^{dp_{k_n}} L_n$. Consequently, sequence $L$ is turned into $\left\{U_1^{dp_{i_1}} U_1^{dp_{k_1}} L_1, U_2^{dp_{i_2}} U_2^{dp_{k_2}} L_2, \cdots, U_n^{dp_{i_n}} U_n^{dp_{k_n}} L_n, \cdots, U_N^{dp_{i_N}} U_N^{dp_{k_N}} L_N\right\}$. Since Bob prepares the logical qubit $L_n$ by himself, he can know its initial state and the measuring basis of the logical qubit $U_n^{dp_{i_n}} U_n^{dp_{k_n}} L_n$. Bob uses the right measuring basis to measure the logical qubit $U_n^{dp_{i_n}} U_n^{dp_{k_n}} L_n$. Then, Bob announces his measurement result of the logical qubit $U_n^{dp_{i_n}} U_n^{dp_{k_n}} L_n$ publicly to Alice, where each announcement needs two classical bits. With the initial state of the logical qubit $L_n$ and his composite unitary operation $U_n^{dp_{i_n}}$, Bob can read out Alice's one-bit secret message $k_n$. As for Alice, according to Bob's announcement on the measurement result of the logical qubit $U_n^{dp_{i_n}} U_n^{dp_{k_n}} L_n$, she can select the right measuring basis to measure the logical qubit $L_n^{'}$ in sequence $L^{'}$. As a result, Alice knows the initial state of the logical qubit $L_n$, as each two adjacent logical qubits $L_n$ and $L_n^{'}$ is prepared in the same state by Bob. Then, With the help of her composite unitary operation $U_n^{dp_{k_n}}$, Alice can also read out Bob's one-bit secret message $i_n$.

Until now, the description of information leakage resistant QD protocol against collective-dephasing noise has been finished. In fact, the basic principle of bidirectional communication in the above protocol is similar to that in Ref.[44]. Different from Ref.[44], the above protocol uses two physical qubits as a logical qubit to prevent the collective-dephasing noise.

In fact, the quantum measurement for decoding in the above protocol can be simplified into the single-photon measurements, instead of the Bell-state measurements. An interesting phenomenon is that a Hadamard operation imposed on each physical qubit can change $\left|+x_{dp}\right\rangle_L$ into $\left|\phi^-\right\rangle_{A_1 A_2} = \frac{1}{\sqrt{2}}\left(|H\rangle_{A_1}|H\rangle_{A_2} - |V\rangle_{A_1}|V\rangle_{A_2}\right)$ and keep $\left|-x_{dp}\right\rangle_L = \left|\psi^-\right\rangle_{A_1 A_2} = \frac{1}{\sqrt{2}}\left(|H\rangle_{A_1}|V\rangle_{A_2} - |V\rangle_{A_1}|H\rangle_{A_2}\right)$ intact, respectively.[20] Obviously, one can easily distinguish the above two Bell states through two single-photon measurements since the parity of two photons is parallel for $\left|\phi^-\right\rangle_{A_1 A_2}$ and antiparallel for $\left|\psi^-\right\rangle_{A_1 A_2}$. Therefore, with a Hadamard operation imposed on each physical qubit first, the two logical qubits $\left|+x_{dp}\right\rangle_L$ and $\left|-x_{dp}\right\rangle_L$ can be easily distinguished through the single-photon measurements, instead of the Bell-state measurements under the $X_L^{dp}$ basis. In this way, both Bob and Alice can directly read out each other's secret messages through the single-photon measurements.

Now the author would like to give a detailed explanation about the bidirectional communication process of the above protocol by using a concrete example. Take the first group $(L_1, L_1^{'})$ as an example. Suppose that $k_1 = 1$ and $i_1 = 0$. Moreover, assume that $L_1$ and $L_1^{'}$ are initially prepared in the state of $\left|0_{dp}\right\rangle_L$. Consequently, after Alice and Bob's encoding, $L_1$ is transformed into $U_1^{dp_0} U_1^{dp_1} L_1 = U_1^{dp_0} U_1^{dp_1} \left|0_{dp}\right\rangle_L = \left|1_{dp}\right\rangle_L$, while $L_1^{'}$ is kept unchanged. Since Bob prepares $L_1$ by himself, he is aware of its initial state and the measuring basis of $U_1^{dp_0} U_1^{dp_1} L_1$. Then, Bob chooses the right measuring basis to measure $U_1^{dp_0} U_1^{dp_1} L_1$, and announces his measurement result publicly. According to the initial state of $L_1$ and his composite unitary operation $U_1^{dp_0}$, Bob can read out that $k_1 = 1$. As for Alice, after receiving Bob's announcement on the measurement result of $U_1^{dp_0} U_1^{dp_1} L_1$, Alice measures $L_1^{'}$ to know the initial state of $L_1$ using the right measuring basis. Then, Alice can also read out that $i_1 = 0$ through her composite unitary operation $U_1^{dp_1}$.

## 2.2 Information leakage resistant QD protocol against collective-rotation noise

The collective-rotation noise in a quantum channel can be modeled as[20,54-59]

$$U_r|H\rangle = \cos\theta|H\rangle + \sin\theta|V\rangle,$$
$$U_r|V\rangle = -\sin\theta|H\rangle + \cos\theta|V\rangle. \tag{5}$$

where $\theta$ is the parameter of collective-rotation noise which fluctuates with time. Apparently, the following two Bell states are immune to this kind of noise:[20,54-59]

$$\left|\phi^{+}\right\rangle_{A_1A_2} = \frac{1}{\sqrt{2}}\left(\left|H\right\rangle_{A_1}\left|H\right\rangle_{A_2} + \left|V\right\rangle_{A_1}\left|V\right\rangle_{A_2}\right), \left|\psi^{-}\right\rangle_{A_1A_2} = \frac{1}{\sqrt{2}}\left(\left|H\right\rangle_{A_1}\left|V\right\rangle_{A_2} - \left|V\right\rangle_{A_1}\left|H\right\rangle_{A_2}\right). \qquad (6)$$

In this case, the logical qubits can be set up by

$$\left|0_r\right\rangle_L = \left|\phi^{+}\right\rangle_{A_1A_2}, \left|1_r\right\rangle_L = \left|\psi^{-}\right\rangle_{A_1A_2}. \qquad (7)$$

The superpositions of these two logical qubits, i.e.,[20,58] $\left|+x_r\right\rangle_L = \frac{1}{\sqrt{2}}\left(\left|0_r\right\rangle_L + \left|1_r\right\rangle_L\right) \equiv \left|\Phi^{+}\right\rangle_{A_1A_2}$ and

$\left|-x_r\right\rangle_L = \frac{1}{\sqrt{2}}\left(\left|0_r\right\rangle_L - \left|1_r\right\rangle_L\right) \equiv \left|\Psi^{-}\right\rangle_{A_1A_2}$, are also immune to this kind of noise. The following four states which span a DFS on this kind of noise are used to construct a quantum channel here:

$$\left|0_r\right\rangle_L = \left|\phi^{+}\right\rangle_{A_1A_2}, \left|1_r\right\rangle_L = \left|\psi^{-}\right\rangle_{A_1A_2}, \left|+x_r\right\rangle_L = \frac{1}{\sqrt{2}}\left(\left|0_r\right\rangle_L + \left|1_r\right\rangle_L\right) \equiv \left|\Phi^{+}\right\rangle_{A_1A_2}, \left|-x_r\right\rangle_L = \frac{1}{\sqrt{2}}\left(\left|0_r\right\rangle_L - \left|1_r\right\rangle_L\right) \equiv \left|\Psi^{-}\right\rangle_{A_1A_2}. \qquad (8)$$

Obviously, $Z_L^r = \left\{\left|0_r\right\rangle_L, \left|1_r\right\rangle_L\right\}$ and $X_L^r = \left\{\left|+x_r\right\rangle_L, \left|-x_r\right\rangle_L\right\}$ are two corresponding measuring bases for the above four states, respectively. Define two composite unitary operations as

$$U^{r_0} = I_{A_1} \otimes I_{A_2}, U^{r_1} = I_{A_1} \otimes \left(-i\sigma_y\right)_{A_2}. \qquad (9)$$

It is distinct that the following relations exist:[20]

$$U^{r_1}\left|0_r\right\rangle_L = \left|1_r\right\rangle_L, U^{r_1}\left|1_r\right\rangle_L = -\left|0_r\right\rangle_L, U^{r_1}\left|+x_r\right\rangle_L = -\left|-x_r\right\rangle_L, U^{r_1}\left|-x_r\right\rangle_L = \left|+x_r\right\rangle_L. \qquad (10)$$

The composite unitary operation $U^{r_1}$ can not alter each of the measuring bases itself but flip its two states inside it. Note that the counterpart of this property in an ideal condition was previously described in Deng and Long's one-time pad QSDC[11].

Suppose that Alice has $N$ bits secret messages $\{k_1, k_2, \cdots, k_N\}$ and Bob has $N$ bits secret messages $\{i_1, i_2, \cdots, i_N\}$, where $k_n, i_n \in \{0,1\}, n \in \{1, 2, \cdots, N\}$. Let each of $U^{r_0}$ and $U^{r_1}$ represent one-bit secret message in such a way that $\{U^{r_0} \to 0, U^{r_1} \to 1\}$. The implementation steps of information leakage resistant QD protocol against collective-rotation noise are extremely similar to those in the case of collective-dephasing noise. As long as the following differences are made, the previous protocol against collective-dephasing noise described in sec.2.1 can be turned into the one against collective-rotation noise:

(1) In Step 1 of the previous protocol, Bob prepares both the logical qubits and the decoy logical qubits randomly in one of the four states $\{\left|0_r\right\rangle_L, \left|1_r\right\rangle_L, \left|+x_r\right\rangle_L, \left|-x_r\right\rangle_L\}$;

(2) In Step 3, Alice uses the composite unitary operation $U_n^{r_{k_n}}$ to encode her one-bit secret message $k_n$. Moreover, Alice uses $U^{r_0}$ or $U^{r_1}$ to encode her checking message on each of $\delta_2$ decoy logical qubits;

(3) In Step 5, Bob uses the composite unitary operation $U_n^{r_{i_n}}$ to encode his one-bit secret message $i_n$. Accordingly, Bob announces his measurement result of the logical qubit $U_n^{r_{i_n}} U_n^{r_{k_n}} L_n$ publicly to Alice.

In fact, the basic principle of bidirectional communication in the above protocol is also similar to that in Ref.[44]. Different from Ref.[44], the above protocol uses two physical qubits as a logical qubit to prevent the collective-rotation noise..

On the other hand, the quantum measurement for decoding in the above protocol can also be simplified into the single-photon measurements, similar to the case of collective-dephasing noise. An interesting phenomenon is that a Hadamard operation imposed on the physical qubit $A_2$ can change $\left|\Phi^{+}\right\rangle_{A_1A_2}$ and $\left|\Psi^{-}\right\rangle_{A_1A_2}$ into $\left|\phi^{-}\right\rangle_{A_1A_2} = \frac{1}{\sqrt{2}}\left(\left|H\right\rangle_{A_1}\left|H\right\rangle_{A_2} - \left|V\right\rangle_{A_1}\left|V\right\rangle_{A_2}\right)$ and $\left|\psi^{+}\right\rangle_{A_1A_2} = \frac{1}{\sqrt{2}}\left(\left|H\right\rangle_{A_1}\left|V\right\rangle_{A_2} + \left|V\right\rangle_{A_1}\left|H\right\rangle_{A_2}\right)$, respectively.[20] Obviously, one can easily distinguish the above two Bell states through two single-photon measurements since $\left|\phi^{-}\right\rangle_{A_1A_2}$ and $\left|\psi^{+}\right\rangle_{A_1A_2}$ have different parities for their own two photons. Therefore, with a Hadamard operation imposed on the physical qubit $A_2$ first, the two logical qubits $\left|\Phi^{+}\right\rangle_{A_1A_2}$ and $\left|\Psi^{-}\right\rangle_{A_1A_2}$ can be easily distinguished through the single-photon measurements, instead of the measurements under the $X_L^r$ basis. On the other hand, $\left|\phi^{+}\right\rangle_{A_1A_2}$ and $\left|\psi^{-}\right\rangle_{A_1A_2}$ also have different parities for their own two photons, and thus one can easily distinguish them through the single-photon measurements, instead of the Bell-state measurements under the $Z_L^r$ basis. In this way, both Bob and Alice can directly read out each other's secret messages through the single-photon measurements.

## 3 Security analysis

Since the basic principles of bidirectional communication in the cases of collective-dephasing noise and collective-rotation noise are identical, without loss of generality, the author takes the first case for example to demonstrate the security analysis.

(1) Analysis of the information leakage problem

The concrete example in section 2.1 is still used to analyze the information leakage problem here. Obviously, during the communication process, $L_1^{'}$ acts as an auxiliary logical qubit used for privately sharing the initial state of $L_1$ between Alice and Bob, making Eve unsure about the initial state of $L_1$. Therefore, although Eve knows the final state of $U_1^{dp_0} U_1^{dp_1} L_1$ from Bob's public announcement, she still cannot get any information about Alice and Bob's secret messages. If she guesses that the initial state of $L_1$ is $\left|0_{dp}\right\rangle_L$, the secret messages will be $\{k_1=0, i_1=1\}$ or $\{k_1=1, i_1=0\}$; if she guesses that the initial state of $L_1$ is $\left|1_{dp}\right\rangle_L$, the secret messages will be $\{k_1=0, i_1=0\}$ or $\{k_1=1, i_1=1\}$. Consequently, there are totally four kinds of uncertainty, containing $-\sum_{i=1}^{4} p_i \log_2 p_i = -4 \times \frac{1}{4} \log_2 \frac{1}{4} = 2$ bit information for Eve from the viewpoint of Shannon's information theory[62], which is equal to the total amount of secret messages from Alice and Bob. Therefore, no information leaks out to Eve. The auxiliary logical qubit $L_1^{'}$ helps directly overcome the information leakage problem in this example.

(2) Analysis of Eve's active attacks

Apparently, during the whole communication, the logical qubit $L_n$ undergoes a round trip, and thus two security checks are needed in total. With respect to the second transmission, Eve can not get any useful information from Alice but disturb the transmission of sequence $L^{''}$ even if she intercepts it. The reason lies in that she is unsure about the preparing bases and the original states of the logical qubits in sequence $L^{''}$. Apparently, the second security check adopts the method of message authentication to estimate whether an Eve is on-line. This security check method can effectively avoid the denial-of-service attack (DoS attack) from Eve. Therefore, the security of the proposed protocol is decided by the first security check. The first security check uses the decoy photon technique [60-61] to guarantee the security of quantum channel, which can be regarded as a variation of the security check method in the BB84 protocol[1]. Concretely speaking, the decoy logical qubits randomly in one of the four states $\left\{\left|0_{dp}\right\rangle_L, \left|1_{dp}\right\rangle_L, \left|+x_{dp}\right\rangle_L, \left|-x_{dp}\right\rangle_L\right\}$ are used to detect the presence of Eve. Now its effectiveness against Eve's several famous active attacks is demonstrated in detail as follows.

① The Trojan horse attacks

There are two kinds of Trojan horse attack strategies, i.e., the invisible photon eavesdropping scheme proposed by Cai[63] and the delay-photon Trojan horse attack[64]. The method presented in Ref.[65] can also be used to resist the Trojan horse attacks here. Concretely speaking, to combat the invisible photon eavesdropping, Alice inserts a filter in front of her devices to filter out the photon signal with an illegitimate wavelength when she receives sequence $S^{'}$ from Bob in Step 2. To combat the delay-photon Trojan horse attack, in Step 2, Alice splits each sampling signal in $\delta_1$ decoy logical photons with a photon number splitter (PNS:50/50) and measures the two signals after the PNS with a proper measuring basis. If the multiphoton rate is unreasonably high, the communication will be halted. Otherwise, the communication is continued. In this way, the above two kinds of Trojan horse attack strategies can be successfully defeated.

② The intercept-resend attack

Eve prepares a fake sequence in advance, which is composed of the logical qubits randomly in one of the four states $\left\{\left|0_{dp}\right\rangle_L, \left|1_{dp}\right\rangle_L, \left|+x_{dp}\right\rangle_L, \left|-x_{dp}\right\rangle_L\right\}$. After intercepting sequence $S^{'}$, Eve substitutes it with her fake sequence and sends the new sequence to Alice. Consequently, Eve's attack can be discovered with a probability of $50\%$, as Alice's measurement results on the fake sequence are not always identical with the genuine ones.

③ The measure-resend attack

After intercepting sequence $S^{'}$, Eve randomly chooses one of the two measuring bases $Z_L^{dp}$ and $X_L^{dp}$ to measure each of its logical qubits. Afterward, she resends it to Alice. Because Eve's measuring bases for decoy logical qubits are not always consistent with their preparing bases form Bob, Eve's attack can be discovered with a probability of $25\%$.

④ The entangle-measure attack

In order to steal partial information, Eve may try to entangle her auxiliary photons $E = \left\{\left|E_1\right\rangle, \left|E_2\right\rangle, \cdots, \left|E_i\right\rangle, \cdots, \left|E_{2N+\delta_1+\delta_2}\right\rangle\right\}$ with the logical qubits in sequence $S^{'}$ through a unitary operation $U_E$. As a result, the system state is transformed into[59]

$$U_E \left|HV\right\rangle \left|E_i\right\rangle = \alpha_{HH}\left|HH\right\rangle\left|e_H e_H\right\rangle + \alpha_{HV}\left|HV\right\rangle\left|e_H e_V\right\rangle + \alpha_{VH}\left|VH\right\rangle\left|e_V e_H\right\rangle + \alpha_{VV}\left|VV\right\rangle\left|e_V e_V\right\rangle,$$

$$U_E \left|VH\right\rangle \left|E_i\right\rangle = \beta_{HH}\left|HH\right\rangle\left|e_H^{'} e_H^{'}\right\rangle + \beta_{HV}\left|HV\right\rangle\left|e_H^{'} e_V^{'}\right\rangle + \beta_{VH}\left|VH\right\rangle\left|e_V^{'} e_H^{'}\right\rangle + \beta_{VV}\left|VV\right\rangle\left|e_V^{'} e_V^{'}\right\rangle,$$

$$U_E|\psi^+\rangle|E_i\rangle = \frac{1}{\sqrt{2}}(U_E|HV\rangle|E_i\rangle + U_E|VH\rangle|E_i\rangle)$$

$$= \frac{1}{2}\begin{pmatrix} |\phi^+\rangle(\alpha_{HH}|e_He_H\rangle + \alpha_{VV}|e_Ve_V\rangle + \beta_{HH}|e_H'e_H'\rangle + \beta_{VV}|e_V'e_V'\rangle) \\ |\phi^-\rangle(\alpha_{HH}|e_He_H\rangle - \alpha_{VV}|e_Ve_V\rangle + \beta_{HH}|e_H'e_H'\rangle - \beta_{VV}|e_V'e_V'\rangle) \\ |\psi^+\rangle(\alpha_{HV}|e_He_V\rangle + \alpha_{VH}|e_Ve_H\rangle + \beta_{HV}|e_H'e_V'\rangle + \beta_{VH}|e_V'e_H'\rangle) \\ |\psi^-\rangle(\alpha_{HV}|e_He_V\rangle - \alpha_{VH}|e_Ve_H\rangle + \beta_{HV}|e_H'e_V'\rangle - \beta_{VH}|e_V'e_H'\rangle) \end{pmatrix},$$

$$U_E|\psi^-\rangle|E_i\rangle = \frac{1}{\sqrt{2}}(U_E|HV\rangle|E_i\rangle - U_E|VH\rangle|E_i\rangle)$$

$$= \frac{1}{2}\begin{pmatrix} |\phi^+\rangle(\alpha_{HH}|e_He_H\rangle + \alpha_{VV}|e_Ve_V\rangle - \beta_{HH}|e_H'e_H'\rangle - \beta_{VV}|e_V'e_V'\rangle) \\ |\phi^-\rangle(\alpha_{HH}|e_He_H\rangle - \alpha_{VV}|e_Ve_V\rangle - \beta_{HH}|e_H'e_H'\rangle + \beta_{VV}|e_V'e_V'\rangle) \\ |\psi^+\rangle(\alpha_{HV}|e_He_V\rangle + \alpha_{VH}|e_Ve_H\rangle - \beta_{HV}|e_H'e_V'\rangle - \beta_{VH}|e_V'e_H'\rangle) \\ |\psi^-\rangle(\alpha_{HV}|e_He_V\rangle - \alpha_{VH}|e_Ve_H\rangle - \beta_{HV}|e_H'e_V'\rangle + \beta_{VH}|e_V'e_H'\rangle) \end{pmatrix},\quad (11)$$

where $|e_He_H\rangle$, $|e_He_V\rangle$, $|e_Ve_H\rangle$ and $|e_Ve_V\rangle$ are Eve's probe states and $|\alpha_{HH}|^2 + |\alpha_{HV}|^2 + |\alpha_{VH}|^2 + |\alpha_{VV}|^2 = |\beta_{HH}|^2 + |\beta_{HV}|^2 + |\beta_{VH}|^2 + |\beta_{VV}|^2 = 1$. In order to pass the first security check, Eve's choice about the above complex numbers should satisfy the following conditions:

$$\begin{cases} \alpha_{HH} = \alpha_{VH} = \alpha_{VV} = 0 \\ \beta_{HH} = \beta_{HV} = \beta_{VV} = 0 \\ \alpha_{HV}|e_He_V\rangle - \alpha_{VH}|e_Ve_H\rangle + \beta_{HV}|e_H'e_V'\rangle - \beta_{VH}|e_V'e_H'\rangle = \vec{0} \\ \alpha_{HV}|e_He_V\rangle + \alpha_{VH}|e_Ve_H\rangle - \beta_{HV}|e_H'e_V'\rangle - \beta_{VH}|e_V'e_H'\rangle = \vec{0} \end{cases} \quad (12)$$

where $\vec{0}$ represents a zero vector. However, it can be derived from formula (12) that $\alpha_{HV}|e_He_V\rangle = \beta_{VH}|e_V'e_H'\rangle$. That is to say, Eve can not distinguish $\alpha_{HV}|e_He_V\rangle$ from $\beta_{VH}|e_V'e_H'\rangle$, making her unable to get partial useful information by measuring her auxiliary photons. On the other hand, if Eve wants to get partial useful information through distinguishing her auxiliary photons (i.e., making $\alpha_{HV}|e_He_V\rangle \neq \beta_{VH}|e_V'e_H'\rangle$), she has to face the danger of being detected by Alice and Bob.[59]

⑤The CE attack

The CE attack, which was firstly proposed by Gao et al. in the analysis of quantum exam [66], always utilizes the relationship between two different qubits or one qubit at different times to extract useful information. Its distinct feature is that it often works on the protocols with entangled states [66-71].

When sequence $S'$ passes by, in order to execute the CE attack, Eve performs the controlled-not (CNOT) operations, $CNOT = |00\rangle\langle 00| + |01\rangle\langle 01| + |11\rangle\langle 10| + |10\rangle\langle 11|$, on the logical qubits in it and her auxiliary photon, where the physical qubits $A_1$ and $A_2$ are the control qubits and $E_i$ is the target qubit. Without loss of generality, assume that the auxiliary photon prepared by Eve is in the state $|H\rangle_{E_i}$. Then, the quantum system evolves into[59]

$$(CNOT_{A_1,E_i})(CNOT_{A_2,E_i})|H\rangle_{A_1}|V\rangle_{A_2} \otimes |H\rangle_{E_i} = |H\rangle_{A_1}|V\rangle_{A_2} \otimes |V\rangle_{E_i},$$

$$(CNOT_{A_1,E_i})(CNOT_{A_2,E_i})|V\rangle_{A_1}|H\rangle_{A_2} \otimes |H\rangle_{E_i} = |V\rangle_{A_1}|H\rangle_{A_2} \otimes |V\rangle_{E_i},$$

$$(CNOT_{A_1,E_i})(CNOT_{A_2,E_i})|\psi^+\rangle_{A_1A_2} \otimes |H\rangle_{E_i} = \frac{1}{\sqrt{2}}(|H\rangle_{A_1}|V\rangle_{A_2} + |V\rangle_{A_1}|H\rangle_{A_2}) \otimes |V\rangle_{E_i} = |\psi^+\rangle_{A_1A_2} \otimes |V\rangle_{E_i},$$

$$(CNOT_{A_1,E_i})(CNOT_{A_2,E_i})|\psi^-\rangle_{A_1A_2} \otimes |H\rangle_{E_i} = \frac{1}{\sqrt{2}}(|H\rangle_{A_1}|V\rangle_{A_2} - |V\rangle_{A_1}|H\rangle_{A_2}) \otimes |V\rangle_{E_i} = |\psi^-\rangle_{A_1A_2} \otimes |V\rangle_{E_i}. \quad (13)$$

Apparently, after the CE attack, the two initial states $|0_{dp}\rangle_L$ and $|1_{dp}\rangle_L$ and the two initial states $|+x_{dp}\rangle_L$ and $|-x_{dp}\rangle_L$ always keep

8unchanged under the $Z_L^{dp}$ basis and the $X_L^{dp}$ basis, respectively. That is to say, Eve's CE attack can escape from the first security check. Fortunately, the initial state $|H\rangle_{E_i}$ is always transformed into $|V\rangle_{E_i}$ after the CE attack. In other words, the final state of the auxiliary photon $E_i$ (i.e., $|V\rangle_{E_i}$) corresponds to four possibilities about the initial state of each logical qubit in sequence $S'$ (i.e., $\{|0_{dp}\rangle_L, |1_{dp}\rangle_L, |+x_{dp}\rangle_L, |-x_{dp}\rangle_L\}$), making Eve unable to know the initial state of each logical qubit in sequence $S'$ exactly by measuring her auxiliary photon $E_i$. Therefore, it can be concluded in this case that although her CE attack cannot be discovered during the first security check, Eve gets nothing useful about the initial states of logical qubits in sequence $S'$.

Based on the above analysis, it can be concluded that the proposed protocol is always secure against Eve's different kinds of active attacks.

## 4 Discussions and conclusions

(1) The information-theoretical efficiency

The information-theoretical efficiency defined by Cabello[3] is $\eta = b_s/(q_t + b_t)$, where $b_s$, $q_t$ and $b_t$ are the expected secret bits received, the qubits used and the classical bits exchanged between Alice and Bob, respectively. In the proposed QD protocol under the case of collective-dephasing noise (collective-rotation noise), without considering the two security checks, each two adjacent logical qubits $L_n$ and $L_n'$ can be used for exchanging Alice's one-bit secret and Bob's one-bit secret with two classical bits consumed for the announcement on the measurement result of the logical qubit $U_n^{dp_{i_n}} U_n^{dp_{k_n}} L_n$ ($U_n^{r_{i_n}} U_n^{r_{k_n}} L_n$). Accordingly, it follows that $b_s = 2$, $q_t = 4$ and $b_t = 2$, making $\eta = \frac{2}{4+2} \times 100\% = 33.3\%$ in both of the two proposed protocols.

(2) Comparisons of previous information leakage resistant QD protocols

It is distinct that all the QD protocols in Refs.[43-48] have the ability to overcome the information leakage problem. However, they are merely designed on the basis of an ideal condition, so they cannot work in a noisy environment. Compared with them, the advantage of the two proposed QD protocols lies in that they can work in a noisy environment.

Apparently, the QD protocols in Ref.[59] are able to not only overcome the information leakage problem but also work in a noisy environment. As a result, with respect to the information leakage resistant property and the anti-noise property, the protocols in Ref.[59] and the two proposed protocols are consistent. Therefore, it is necessary to compare the two proposed protocols with them in detail. Their comparisons are concentrated on three aspects containing the quantum resource, the quantum measurement and the information-theoretical efficiency. As to the quantum resource, two-photon states are consumed in the proposed two protocols while four-photon states are consumed in the protocols of Ref.[59]. Because the preparation of two-photon states is more convenient than that of four-photon states, the two proposed protocols exceed the protocols of Ref.[59] in the quantum resource. As to the quantum measurement, for decoding, single-photon measurements are needed in the two proposed protocols while Bell-state measurements are needed in the protocols of Ref.[59]. Because single-photon measurements are easier to perform than Bell-state measurements, the two proposed protocols exceed the protocols of Ref.[59] in the quantum measurement. In addition, in the protocols of Ref.[59], each product state $IS \otimes MS$ can be used for exchanging Alice's one-bit secret and Bob's one-bit secret with one classical bit consumed for the announcement on the measurement result $M_{AB}$. As a result, the information-theoretical efficiency of the protocols in Ref.[59] is $\eta = \frac{2}{4+1} \times 100\% = 40\%$. Therefore, compared with the protocols of Ref.[59], the two proposed protocols have a poorer performance on the information-theoretical efficiency.

To sum up, in this paper, two information leakage resistant QD protocols over a collective-noise channel are proposed. DFS is used to erase the influence from two kinds of collective noise, i.e., collective-dephasing noise and collective-rotation noise, where each logical qubit is composed of two physical qubits and free from noise. In each of the two proposed protocols, the secret messages are encoded on the initial logical qubits via two composite unitary operations. Moreover, the single-photon measurements rather than the Bell-state measurements or the more complicated measurements are needed for decoding, making the two proposed protocols easier to implement. The initial state of each logical qubit is privately shared between the two authenticated users through the direct transmission of its auxiliary counterpart. Consequently, the information leakage problem is avoided in the two proposed protocols. Moreover, the detailed security analysis also shows that Eve's several famous active attacks can be effectively overcome, such as the Trojan horse attack, the intercept-resend attack, the measure-resend attack, the entangle-measure attack and the CE attack.

**Acknowledgements**

Funding by the National Natural Science Foundation of China (Grant No.11375152), and the Natural Science Foundation of Zhejiang Province (Grant No.LQ12F02012) is gratefully acknowledged.